\documentclass[twocolumn,preprintnumbers,prl,amsmath,amssymb,superscriptaddress,showpacs]{revtex4}
\usepackage{epsfig}
\usepackage{graphicx}
\usepackage{color}

\begin{document}

\title{Provably Secure Experimental Quantum Bit-String Generation}

\author{L.P. Lamoureux}
\affiliation{Laboratoire d'Information Quantique and 
Quantum Information and Communication, 
 CP 165, Universit\'e Libre de Bruxelles, Avenue F. D.
Roosevelt 50, 1050
Bruxelles, Belgium}

\author{E. Brainis}
\affiliation{Optique et Acoustique,
CP 194/5, Universit\'{e} Libre de Bruxelles, Avenue F. D.
Roosevelt 50, 1050 Bruxelles, Belgium}

\author{D. Amans}
\affiliation{Optique et Acoustique,
CP 194/5, Universit\'{e} Libre de Bruxelles, Avenue F. D.
Roosevelt 50, 1050 Bruxelles, Belgium}

\author{J. Barrett}
\affiliation{Laboratoire d'Information Quantique and 
Quantum Information and Communication, 
 CP 165, Universit\'e Libre de Bruxelles, Avenue F. D.
Roosevelt 50, 1050
Bruxelles, Belgium}

\author{S. Massar}
\affiliation{Laboratoire d'Information Quantique and 
Quantum Information and Communication, 
 CP 165, Universit\'e Libre de Bruxelles, Avenue F. D.
Roosevelt 50, 1050
Bruxelles, Belgium}

\date{\today}

\begin{abstract}
Coin tossing is a cryptographic task in which two parties who do
not trust each other aim to generate a common random bit. Using
classical communication this is impossible, but non trivial coin
tossing is possible using quantum communication. Here we consider
the case when the parties do not want to toss a single coin, but
many. This is called bit string generation. We report the
experimental generation of strings of coins which are provably
more random than achievable using classical communication. The
experiment is based on the ``plug and play'' scheme developed for
quantum cryptography, and therefore well suited for long distance
quantum communication.
\end{abstract}

\maketitle

Coin tossing is a cryptographic task, introduced by Blum
\cite{Blum}, in which two parties who do not trust one another aim
to generate a common random bit. Coin tossing is an important
primitive that can be used in the design of other two-party
protocols such as mental poker and mail certification and it could
even form the basis of a scheme for bit commitment that is
computationally secure against quantum attacks \cite{kentuses}.
Classically, coin tossing is impossible without computational
assumptions: at least one of the parties can in principle always
cheat and fix the outcome. Using quantum communication, however,
non-trivial coin tossing is possible
\cite{spekkensrudolphdegrees,spekkensrudolphcheatsensitive,ambainislowerbound,%
ambainisnewprotocol,mochon}. 
In many applications, the parties do not
want to generate a single coin, but many. This is called
bit-string generation \cite{Kent,BM,BM2}. Here we report on an
experimental implementation of bit-string generation based on the
``plug and play'' scheme developed for Quantum Key Distribution
(QKD) in optical fibers at telecommunication wavelengths
\cite{PlugandPlay}. Using the theoretical analysis of \cite{BM2}
we are able to show that the bit strings generated in our
experiment achieve a level of randomness impossible classically.
This is the first demonstration of a fundamental new concept:
namely the possibility of generating random coins with an
adversary who is limited only by the laws of physics. 

The present work focuses on bit string generation rather than the
tossing of a single coin for two reasons. First it is shown in
\cite{BM2} that in principle arbitrarily high levels of randomness per
bit can be obtained for bit string generation whereas this is not the
case for coin tossing\cite{lochau,Kitaev}. Hence bit string generation
is more promising from the point of view of applications. Second,
present experimental limitations (mainly detector noise and
inefficiency) seem to preclude tossing a single coin with a level of
randomness higher than what is possible classicaly. This difficulty is
illustrated by another experiment
which recently realized some aspects of coin tossing
\cite{Z}, but for which it was impossible to prove that 
a level of randomness impossible classically was achieved.

We begin by reviewing security conditions for the generation of $n$
random bits. The outcome of the protocol is either a string of bits
$\vec{x} \in \{0,1\}^n$ or one of the parties aborts, in which case
we write $\vec{x}=\perp$. The protocol is {\em correct} if when both
parties are honest, the probability of aborting is small and all the
coins are fair. Mathematically we express this as
\begin{eqnarray}
\label{bscorrect}
\forall \vec{c} \in \{0,1\}^n \quad \mathrm{P}(\vec{x} = \vec{c})
&=& (1 - \delta_n) / 2^n,\nonumber\\
\mathrm{P}(\vec{x} = \perp)&=& \delta_n .
\end{eqnarray}
It is 
necessary to include the parameter $\delta_n$ because of experimental
imperfections which induce a non-zero probability of the protocol
aborting even if both parties are honest. In the protocol we use
$\delta_n$ decreases to zero exponentially fast with $n$ and can
be neglected.

We shall use two security conditions. The first, called the ``average
bias'', describes the degree
of randomness of individual bits of the string. Formally we define
 the upper bound  $\overline{\epsilon_{A(B)}}$ on the average bias 
when
Alice (Bob) is dishonest and the other party is honest as:
\begin{eqnarray}
\forall S_A \forall \vec{c} \in \{0,1\}^n\quad {1 \over
n}\sum_{i=1}^n \mathrm{P}^{S_A H_B}(x_i = c_i) \leq {1 \over 2} +
\overline{\epsilon_A},\nonumber\\
\forall S_B \forall \vec{c} \in \{0,1\}^n\quad {1 \over
n}\sum_{i=1}^n \mathrm{P}^{H_A S_B}(x_i = c_i) \leq {1 \over 2} +
\overline{\epsilon_B},
\end{eqnarray}
where we denote a general strategy of Alice (Bob) by $S_A$ ($S_B$),
and the honest strategy defined by the protocol as $H_A$ ($H_B$).
Classically, when $\delta_n=0$, one has
$\overline{\epsilon_A}+\overline{\epsilon_B} \geq 1/2$ \cite{BM2}.
(When $\delta_n\neq 0$ the
classical bound becomes 
$\overline{\epsilon_A}+\overline{\epsilon_B} \geq 1/2 - 2
\delta_n$.) 

The second security condition measures the degree of randomness of the
string taken as a whole. We define $H_{A(B)}$ as the entropy of the
string if Alice (Bob) is dishonest and the other party is honest.
 In \cite{BM2}, bounds on
the entropy are derived for our protocol assuming general cheating.
However the corresponding classical bound is not known, although it is  
conjectured in \cite{BM2} to be of the form $H_A + H_B \leq n +
o(n)$. We refer to \cite{BM2} for a more detailed discussion of
security conditions and for formal definitions of $H_{A(B)}$.

The protocol we shall use, inspired by that of \cite{BM,BM2} is as
follows. Choose a security parameter $0<\kappa<1$.
\begin{enumerate}
\item
For $i=1$ to $n$.
\item
Alice chooses a random bit $a_i$. If $a_i=0$, she prepares a
coherent state of the electromagnetic field with amplitude
$\alpha$: $\psi_0=|\alpha\rangle$. If $a_i=1$, she prepares a
coherent state with amplitude $-\alpha$: $\psi_1 =
|-\alpha\rangle$. She sends the coherent state $\psi_{a_i}$ to
Bob. After receiving the quantum state from Alice,
Bob chooses a random bit $b_i$. Bob tells Alice the
value of $b_i$.
\item
After learning the value of $b_i$, Alice reveals the value of
$a_i$ to Bob.
\item
Bob now verifies whether the state Alice sent him is indeed the
coherent state $|(-1)^{a_i}\alpha \rangle$. He does this by using
a Local Oscillator (LO) to carry out the displacement
${\cal D}(-(-1)^{a_i}\alpha )$. If Alice was honest, the displaced
state should be the vacuum state. Bob checks that this is the case
by sending the state onto a single photon detector. If the detector
clicks, Bob sets $k_i=1$. If the detector does not click, Bob sets $k_i=0$.
\item
Next $i$.
\item
If $\frac{1}{n}\sum_i k_i > \kappa$, Bob aborts. Otherwise
the output of the protocol is the bit string $x_i = (a_i + b_i) {\rm \ mod}\ 2$.
\end{enumerate}

When Bob is dishonest his best strategy is to measure the state sent to him by
Alice as soon as he receives it (i.e., before carrying out step 3 above). One
easily shows, see \cite{BM2}, that
\begin{equation}\label{EBB}
\overline{\epsilon_B} \leq \frac{\sin \theta}{2},
\quad\text{where}\quad \cos \theta = |\langle
\psi_0|\psi_1\rangle| = e^{-2 |\alpha|^2}\ .
\end{equation}

If Alice is dishonest she may not send Bob the state $\psi_{a_i}$ but
an arbitrary 
state $\rho$. In general she may prepare an entangled state, keeping
half of it and 
sending the other half to Bob. Furthermore, she may correlate and even
entangle her 
strategy over different runs. In \cite{BM2}, however, it is shown that
strategies 
correlated over different runs cannot help Alice for large $n$. A bound on
$\overline{\epsilon_A}$ is proven that depends on the average value of
the fidelity 
$f_i=\langle \psi_{a_i}|\rho|\psi_{a_i}\rangle$, as estimated by Bob. Since the
probability that Bob's detector clicks (assuming his detector is
perfect) is related 
to the fidelity by $P(k_i=1)\geq 1 - f_i$, the result of \cite{BM2}
then implies that, 
assuming large $n$, the bias if Alice is dishonest is bounded by
$\overline {\epsilon_A}\leq {\cal F}(\kappa)$, where
${\cal F}(x)=\frac{\sqrt{x}}{\sqrt{2}\sin^2 \theta} +
\frac{x}{\sin^2\theta}$. Below 
we show how this relation must be modified to take into account
imperfections in Bob's 
measuring apparatus.

Note that due to such imperfections, Bob's detector may click
even if Alice is honest. Alice and Bob should choose $\kappa$ such
that it is larger than the expected number of clicks if both parties
are honest. When this is the case, the probability $\delta_n$ that the
protocol aborts 
if both parties are honest decreases exponentially fast to zero and
the protocol is  
correct.

\begin{figure*}
\centerline{\includegraphics[width=0.75\linewidth]{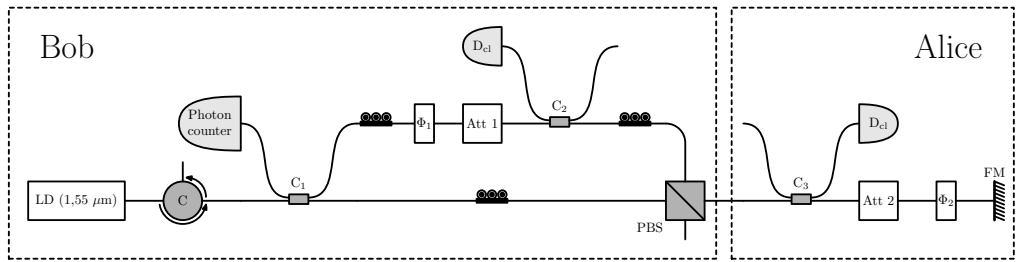}}
\caption{Optical setup. LD: Laser Diode, $\text{C}_i$ (i=1,2,3):
Coupler, Att: Attenuator, $\Phi$: phase modulator, FM: Faraday
Mirror, $\text{D}_{\text{cl}}$: classical detector.} \label{fig1}
\end{figure*}

Our experimental setup, depicted in Fig.~\ref{fig1}, is based on
the plug and play system developed for long distance QKD
\cite{PlugandPlay}.
The advantage of the plug and play system is
that it constitutes an all-fiber (standard SMF-28),
automatically balanced interferometer, and hence is well suited to
long distance quantum communication. However the plug and play system
has a number of specific features which must be carefully taken into account.

{\em Bob to Alice and Bob's cheating.}
Each round of the protocol begins with Bob producing a short
(20ns) intense (25mW) laser pulse at $\lambda=1.55\mu$m. The pulse
is split in two by the 50/50 coupler $\mbox{C}_1$. The two pulses
acquire a relative time delay of 100ns and then impinge with
orthogonal polarization on a Polarizing Beam Splitter (PBS)
whereupon they are sent to Alice. Between $\mbox{C}_1$ and the
PBS, along the long path, are an attenuator, a 99/1 coupler
$\mbox{C}_2$ and a phase modulator. The role of these elements
will be explained later. The relative attenuation of the two
pulses is $A\simeq 45$dB. The first pulse to reach Alice is
intense and contains $N_0\simeq 10^9$ photons. This pulse will
play the role of LO. The second pulse to reach Alice is attenuated
and contains $A N_0$ photons. The second pulse will play the role
of signal.

Upon receiving the pulses, Alice measures the intensity of the
signal pulse (using the 80/20 coupler $\mbox{C}_3$ and a classical
detector $\text{D}_{\text{cl}}$) and attenuates both pulses. The
two pulses are reflected by the Faraday mirror and travel back to
Bob. The total attenuation at Alice's site is $A'\simeq 50$dB.
Thus the two pulses now contain $A' N_0$ and $AA'N_0$ photons
respectively. In particular the signal pulse now contains only a
few photons ($AA'N_0 = |\alpha|^2= O(1)$). Alice also adds a phase
$\phi_A = a_i \pi $ to the signal pulse, thereby encoding the
value of her bit $a_i$.

The fact that Bob provides Alice with the signal state seems to
provide him with some simple cheating strategies. For instance he
could provide Alice with a signal state that is squeezed
in phase in order to decrease the
overlap between $|\psi_0\rangle$ and $|\psi_1\rangle$. This
apparently allows him to discriminate much better
$|\psi_0\rangle$ from $|\psi_1\rangle$ and hence the value of
$a_i$. The role of the attenuation is to prevent this kind of
cheating. Indeed under strong attenuation any quantum state tends
towards a mixture of coherent states. 

To show this we describe the state by its generalized Wigner
function $W(q,p,s)$. We recall that $W(s=-1)$ is the $Q$ function
which is always positive, $W(s=0)$ is the Wigner function, and
$W(s=+1)$ is the $P$ function. If the $P$ function is positive,
then the state is a mixture of coherent states. Under attenuation
by $A$ we have (see \cite{L}): $ W^{out}(q,p,s)= \frac{1}{A}
W^{in}(\frac{q}{\sqrt{A}},\frac{p}{\sqrt{A}}, \frac{s+A-1}{A}) $
which implies that $W^{out}(s=1-2A)$ is positive. This expresses
the fact that for $A\rightarrow 0$ one tends towards a positive
$P$ function. This result can be made more quantitative by
supposing that after attenuation we add a small amount of Gaussian
noise with mean number of chaotic photons $n$. This affects the
Thus attenuation followed by addition of chaotic photons yields the
transformation
$
W^{out}(q,p,s)= \frac{1}{A}
W^{in}(\frac{q}{\sqrt{A}},\frac{p}{\sqrt{A}}, \frac{s-2n +A-1}{A})
$
and in particular if $n=A$ we have
$
W^{out}(q,p,s=+1)= \frac{1}{A}
W^{in}(\frac{q}{\sqrt{A}},\frac{p}{\sqrt{A}}, s=-1)
$,
i.e. the output $P$ function is positive since it is given in terms of the
input $Q$ function. Thus after strong attenuation, say $A= 10^{-3}$, a quantum
state is very well approximated by a mixture of coherent states since
a very small amount of Gaussian noise with mean number of chaotic photons
$n=10^{-3}$ transforms the state into a mixture of coherent states.

Another simple cheating strategy is for Bob to increase the
intensity of the signal state since it is then much easier for him
to estimate the phase $\phi_A$. The role of the classical
intensity measurement is to ensure that the signal state Alice
sends back is not too intense. In fact it is impossible for Bob to
exploit the fact that he provides Alice with the light pulse which
will become the signal state, since by measuring the intensity of
the pulse Bob sends her and then attenuating it, Alice ensures
that she sends back to Bob a coherent state of known intensity.

Note that the classical intensity measurement of Alice will be
affected by noise because 
$A N_0$ is close to the sensitivity limit of Alice's detector. We
circumvent this 
technical problem by letting Alice carry out statistical tests on the
$n$ intensity 
measurements (one for each round of the protocol). More precisely she
checks whether 
the distribution of measured intensities is consistent with the
Gaussian distribution 
she expects from instrumental noise. If it is she has a precise
estimate of $|\alpha|^2$, 
and hence of $\overline{\epsilon_B}$ through eq.~(\ref{EBB}). If it is
not she aborts.

{\em From Alice to Bob and Alice's cheating.}
Upon receiving the two pulses from Alice, Bob uses coupler
$\mbox{C}_2$ to measure the intensity of the LO, attenuates it by
$A$, and adds a phase $e^{i\phi_B}$, with $\phi_B=a_i\pi$. Note
that by measuring the intensity of the LO state provided by Alice
and then attenuating it, Bob ensures that the LO he uses is a
coherent state (or a mixture of coherent states) of known
intensity $|\beta|^2$. (The argument is exactly the same as that
given above in the case of Alice).

\begin{figure}
\centerline{\includegraphics[width=0.9\linewidth]{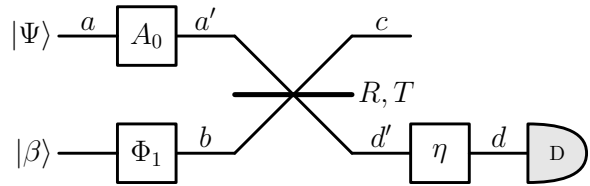}}
\caption{Optical setup equivalent to Bob's measurement, including
its imperfections.} \label{fig2}
\end{figure}

Let us consider the two states that interfere at coupler
$\mbox{C}_1$. On the one hand there is the LO which as we have
just argued is a coherent state of known intensity $|\beta|^2$. On
the other hand there is the signal state. The signal state travels
through the PBS where it gets attenuated by $A_0$. It then
interferes with the LO at coupler $\mbox{C}_1$. This coupler has
transmission and reflection coefficients $T$ and $R$ (both are
approximately 50\%). Finally one of the outputs of the coupler is
sent to a single photon detector (id Quantique) with efficiency
$\eta$. In our experiment $A_0 T=4.3$dB and $\eta=10.5\%$. We can
therefore model the whole of Bob's detection system by the scheme
depicted in Fig.~\ref{fig2}.
It is composed of the LO (a coherent state of amplitude $\beta$),
the signal state $\Psi$, the attenuator $A_0$, a beam splitter
with transmission and reflection coefficients $T$ and $R$. The
imperfect detector is modeled by an attenuation of $\eta$ followed
by a perfect detector.

Let us denote by $\alpha$ the amplitude of the coherent state that
would give rise to destructive interference at the single photon
detector. It satisfies $\alpha \sqrt{A_0 T}+ i \beta \sqrt{R} = 0$.
When $a_i=0$, the state Alice should send if she is honest is the
coherent state $|\alpha\rangle$. (If $a_i=1$ she should send the state
$|-\alpha\rangle$. By using the phase modulator Bob can cancel this phase).
But if Alice is dishonest she will send another state
$|\Psi\rangle$. We expand $|\Psi\rangle$ in the basis of displaced
Fock states
$
|\Psi\rangle
=D_a(\alpha)\sum_n c_n |n\rangle
$
where $D_a(\alpha)$ is the displacement operator acting on mode $a$,
ie.
$
D_a(\alpha)aD_a(\alpha)^\dagger = a - \alpha
$, and
$
|n\rangle =
(a^\dagger)^n/\sqrt{n!}|0\rangle
$ are the Fock states.
The fidelity of the state sent by Alice is thus $f=|\langle
\alpha|\Psi\rangle|^2 = |c_0|^2$.

We model the effect of the attenuation by the transformation $a
\to \sqrt{A_0}a'+ \sqrt{1-A_0}e_1 $ where $e_1$ is a mode of the
environment; the effect of the BS by the transformations $ a'\to
\sqrt{T}d' - i \sqrt{R}c $, $ b\to \sqrt{T}c - i \sqrt{R}d' $; and
the effect of the detector inefficiency by $ d'\to \sqrt{\eta}d +
\sqrt{1-\eta}e_2 $ where $e_2$ is another mode of the environment
(the modes $a,a',b,c,d',d$ are all described in the figure). One
then finds that the state just before entering the single photon
detector is $
D_c(\gamma)\sum_n \frac{c_n}{\sqrt{n!}} [\sqrt{A_0T\eta}d^\dagger
+ i\sqrt{A_0R}c^\dagger 
+\sqrt{1-A_0}e_1^\dagger + \sqrt{(1-\eta)T A_0}e_2^\dagger]^n
|0\rangle
$ where $ \gamma = \beta\sqrt{T} + i \alpha \sqrt{A_0R} $. From
this one easily computes that the probability that the detector
does not register a single click is
\begin{equation}
P(no\ click)= \sum_{n=0}^\infty |c_n|^2 (1-A_0T\eta)^n.
\end{equation}
The probability of registering a click is thus bounded by
$P(click)\geq (1-|c_0|^2)A_0T\eta$. Thus the number of clicks on
Bob's detector divided by $A_0T\eta$ gives a bound on the fidelity
$|c_0|^2$.

A final inefficiency that must be taken into account is that Bob's
detector will have a non-zero dark count rate $\kappa_{dark}= 9 \
10^{-4}$. Putting 
all this together we deduce the bound on the average bias if Alice is
dishonest: 
\begin{equation}
\overline{\epsilon_A} \leq {\cal F}(\frac{\kappa - \kappa_{dark}}{A_0T\eta})
\label{EAA}
\end{equation}
Note that this bound on $\overline{\epsilon_A}$ is given entirely by parameters
which can be measured by Bob.

\begin{figure}
\centerline{\includegraphics[width=0.9\linewidth]{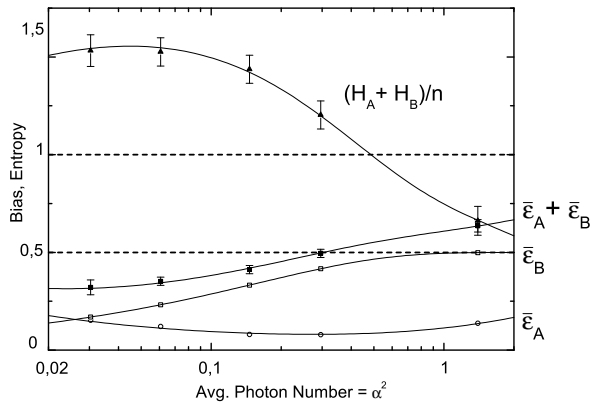}}
\caption{Measured bounds on average bias and on entropy of bit
strings for different values of the average photon number
$\alpha^2$. Open squares: bounds on $\overline{\epsilon_B}$
obtained using eq. (\ref{EBB}); open circles: bounds on
$\overline{\epsilon_A}$ obtained using eq. (\ref{EAA}): filled
squares: bounds on $\overline{\epsilon_A} +
\overline{\epsilon_B}$. Classically the sum is always greater than
$1/2$. The bit strings are clearly more random than is allowed by
the best classical protocol.  The same expressions which give bounds on
$\overline{\epsilon_A}$, $\overline{\epsilon_B}$ also give lower
bounds on the entropies $H_{A(B)}$ of the bit string if Alice
(Bob) is dishonest (see \cite{BM2}). Filled triangles: bounds on the
entropy per bit $(H_A + H_B)/n$. It is conjectured in \cite{BM2}
that for any classical protocol $(H_A + H_B)/n$ is bounded by $1$
for large $n$. The experimental points are clearly above this
bound. 
The error bars for $\overline{\epsilon_A} +
\overline{\epsilon_B}$ and $(H_A + H_B)/n$ describe 
systematic errors arising from 
incorrect callibration of detector efficiency $\eta$ 
and incorrect estimation of $\alpha^2$.
The plotted curves 
are theoretical predictions based on the observed optical visibility of
$96.5\%$. For $\overline{\epsilon_B}$ the curve is given by
eq. (\ref{EBB}) and for $\overline{\epsilon_A}$ it is given by
eq. (\ref{EAA}) using the fact that for small $\alpha^2$, $(\kappa -
\kappa_{dark}) / A_0 T \eta\simeq (1 - V) \alpha^2 /2$.
} \label{fig3}
\end{figure}

Using this protocol, and taking into account experimental
imperfections as described below, a typical run of our experiment
generates $10^7$ coins. Some results for different values of
$|\alpha|^2$ are presented in Fig.~\ref{fig3}.
For instance when $|\alpha|^2=0.03$, we obtained
$\overline{\epsilon_A}+\overline{\epsilon_B}=0.32\pm 0.04$, which is
significantly better than the classical bound
$\overline{\epsilon_A}+\overline{\epsilon_B}\geq 1/2$.

An important property of this protocol and of its experimental
implementation is that we do not have to make any hypothesis about
the Hilbert space Alice or Bob use if they are dishonest -for instance
it is not necessary to restrict them to the single photon
subspace-, nor do we have to make any hypothesis about the kind of
technology they can use if they are dishonest. Thus the randomness
of the bit string when one of the parties is dishonest
is guaranteed by the laws of physics.

\acknowledgments
The authors thank Jarom\'{\i}r Fiur\'{a}\u{s}ek for helpful
discussions. They acknowledge financial support from the Action de
Recherche Concert{\'e}e de la Communaut\'e Fran{\c{c}}aise de Belgique,
from the IUAP program of the Belgian Federal Governement under grant V-18
and from the European Union through project RESQ IST-2001-37559.

\end{document}